\documentclass[10pt,english,journal]{IEEEtran}
\usepackage[T1]{fontenc}
\usepackage[latin9]{inputenc}
\usepackage{geometry}
\usepackage{amsmath}
\usepackage{amssymb}
\usepackage{graphicx}
\usepackage{esint}
\usepackage{epstopdf}

\makeatletter

\newcommand{\noun}[1]{\textsc{#1}}

\renewcommand\[{\begin{equation}}
\renewcommand\]{\end{equation}} 
\newcounter{MYtempeqncnt}
\usepackage{listings}
\usepackage{fancyvrb}
\usepackage{framed}

\usepackage{courier}
\usepackage[usenames,dvipsnames]{color}

\definecolor{dkgreen}{rgb}{0,0.3,0}
\definecolor{gray}{rgb}{0.5,0.5,0.5}

\makeatother

\usepackage{babel}
\begin{document}

\title{Performance Analysis of Project-and-Forward Relaying in~Mixed MIMO-Pinhole and Rayleigh Dual-Hop Channel}

\vskip 0.4cm
\author{Hatim Chergui,~\IEEEmembership{Member,~IEEE}, Mustapha Benjillali,~\IEEEmembership{Senior Member,~IEEE},\\ and~Samir Saoudi,~\IEEEmembership{Senior Member,~IEEE}}

\maketitle
\thispagestyle{empty}

\begin{abstract}
In this letter, we present an end-to-end performance analysis
of dual-hop project-and-forward relaying in a realistic scenario, where the source-relay and
the relay-destination links are experiencing MIMO-pinhole and Rayleigh channel conditions, respectively. We derive the probability density
function of both the relay post-processing and the end-to-end signal-to-noise
ratios, and the obtained expressions are used to derive the outage
probability of the analyzed system as well as its end-to-end ergodic capacity in terms of generalized functions. Applying then the residue theory to Mellin-Barnes integrals, we infer the system asymptotic behavior for different channel parameters. As the bivariate Meijer-G function is involved in the analysis, we propose a new and fast MATLAB implementation enabling an automated definition of the complex integration contour. Extensive Monte-Carlo simulations are invoked to corroborate the analytical results.
\end{abstract}


\section{Introduction}

\PARstart{O}{ne} detrimental situation to MIMO communication benefits is the pinhole effect that usually arises when the transmit-receive range is much larger than the radii of local scatterers in both sides. In that case, the fading energy propagates through a very thin air pipe, called a {\it pinhole} (or keyhole), reducing the MIMO channel to a rank-one matrix~\cite{MIMO_Outdoor}.

In downlink dual-hop multi-antenna relaying systems, the pinhole scenario may practically surface in either hops. Hence, in rich-scattering dense urban fixed deployments, a carefully planned relay location ensures a full-rank source-relay channel; while the relay-destination link may endure the pinhole effect for user equipments (UEs) experiencing poor scattering situations. Conversely, in suburban and rural areas with green-field deployment, the donor eNodeB and the relay are separated by a large distance in a line of sight (LOS) environment such that the source-relay channel has only one degree of freedom \cite{MIMO_Outdoor, Long_Distance}. On the other hand, the fact that the relay is close to the target destination---e.g., a village presenting rich scattering and short ranges to the end UEs---leads to a full-rank Rayleigh relay-destination link. This scenario is also applicable to moving relay nodes (MRNs) in high speed vehicles \cite{MRN}, where the large rural eNodeB LOS coverage and the rare handover events induce large eNodeB-relay distances, and therefore the pinhole effect, while the rich-scattering indoor structure of the vehicle (like trains for instance) and the small relay-UEs ranges yield a Rayleigh propagation.%

An inherent limitation in amplify-and-forward (AF) relaying systems is the so-called noise amplification and propagation that becomes even worse when the number of relay antennas increases, as the corresponding relayed noises accumulate at each of the destination receive antennas; the end-to-end SNR, and therefore the performance, are consequently degraded. To sidestep this drawback, a variant of AF relaying, termed ``project-and-forward'' (PF), has been introduced in \cite{Multihop_DMT}, and consists on optimizing the number of active antennas at the relay by forwarding the degrees-of-freedom (DoF) of the received signal---yield by an orthogonal projection---instead of the signal itself. Only as few relay antennas as the rank of the source-relay MIMO channel are used, i.e., a single antenna in the unit-rank case.

While the mixed full-rank/pinhole MIMO channel has been widely studied in the literature, especially for AF-based setups (cf. \cite{Keyhole1,Keyhole2} and references therein), the MIMO-pinhole/Rayleigh channel has been rarely addressed and, to the best of our knowledge, never for the PF scheme that, in addition, turns out to be very opportune in such environments.

In this letter, we present a novel end-to-end performance analysis of dual-hop PF systems over the mixed MIMO-pinhole/Rayleigh relay channel. We derive exact expressions for the probability density functions (PDFs) of both the first hop and the end-to-end SNRs, which are then used to infer the outage probability as well as the ergodic capacity whose formula is provided in terms of the bivariate Meijer G-function \cite{Bivariate_Meijer_G_def}. The asymptotic behavior is then derived using the residue theory. While the Meijer G-function \cite[Eq. (9.301)]{Table_of_Integrals} is a built-in routine in prevalent computing softwares, the bivariate Meijer G-function is available only in MATHEMATICA with no general contour definition \cite{BER_Bivariate_Meijer_G}. We therefore develop a fast MATLAB code with automated integration contour for this generalized function as a secondary contribution of this work.

In the sequel, the superscript $^{\mathrm{H}}$ denotes the Hermitian transpose, $\left\Vert \cdot\right\Vert _{\mathrm{F}}$ and $\mathrm{Res}\left[\phi,p\right]$ represent the Frobenius norm and the residue of function $\phi$ at pole $p$. $\Gamma\left(\cdot\right)$, $\psi^{(0)}\left(\cdot\right)$, and $K_{\nu}\left(\cdot\right)$
stand for the Gamma function, the digamma function, and the $\nu^{th}$-order modified Bessel function of the second kind, respectively. $\mathrm{G}_{\cdot,\cdot}^{\cdot,\cdot}\left(\cdot\mid\cdot\right)$ is the Meijer G-function, and $\mathrm{G}_{\cdot,\cdot:\cdot,\cdot:\cdot,\cdot}^{\cdot,\cdot:\cdot,\cdot:\cdot,\cdot}\left(\cdot,\cdot\mid\cdot\mid\cdot\mid\cdot\right)$ is the bivariate Meijer G-function.

\section{System Model}

\subsection{Channel Description}

We consider a half-duplex dual-hop multi-antenna cooperative transmission
where an $n_{\mathrm{s}}$-antennas source
is connected to a single antenna destination through an
$n_{\mathrm{r}}$-antennas relay ($n_{\mathrm{s}},n_{\mathrm{r}}{>}1$). The communication between each couple of nodes, $\mathrm{i}\hspace{-1mm}\in\hspace{-1mm}\left\{ \mathrm{s,r}\right\} $
and $\mathrm{i'}\hspace{-1mm}\in\hspace{-1mm}\left\{ \mathrm{r,d}\right\} $,
takes place over an independent wireless link $\mathrm{i-i'}$ experiencing
an average propagation loss $\alpha_{\mathrm{ii'}}$. The corresponding small scale
fading effects are represented by
\begin{itemize}
\item A MIMO-pinhole channel matrix $\mathbf{H}_{\mathrm{sr}}$ that is modelled as an outer product of two independent and uncorrelated
Rayleigh fading vectors $\mathbf{g}_{\mathrm{s}}\in\mathbb{C}^{n_{\mathrm{s}}\times1}$
and $\mathbf{g}_{\mathrm{r}}\in\mathbb{C}^{n_{\mathrm{r}}\times1}$,
i.e.,
\[
\mathbf{H}_{\mathrm{sr}}=\mathbf{g}_{\mathrm{r}}\mathbf{g}_{\mathrm{s}}^{\mathrm{H}}\in\mathbb{C}^{n_{\mathrm{r}}\times n_{\mathrm{s}}}.
\]

\item An independent standard complex Gaussian vector 
$\mathbf{h}_{\mathrm{rd}}$ whose coefficients $\left\{ h_{\mathrm{rd}}^{n,n'}\right\} $ 
are consequently Rayleigh distributed.
\end{itemize}
Both relay and destination received signals are corrupted by additive
white Gaussian noise (AWGN) vectors $\mathbf{w}_{\mathrm{r}}\sim\mathcal{N}\left(\mathbf{0}_{n_{\mathrm{r}}\times1},\sigma^{2}\mathbf{I}_{n_{\mathrm{r}}}\right)$
and $\mathbf{w}_{\mathrm{d}}\sim\mathcal{N}\left(\mathbf{0}_{n_{\mathrm{d}}\times1},\sigma^{2}\mathbf{I}_{n_{\mathrm{d}}}\right)$,
respectively. The corresponding average SNRs per hop are $\overline{\gamma}_{\mathrm{sr}}=\alpha_{\mathrm{sr}}^{2}/\sigma^{2}$
and $\overline{\gamma}_{\mathrm{rd}}=\alpha_{\mathrm{rd}}^{2}/\sigma^{2}$.

\subsection{Project-and-Forward Relaying}

Let $\mathbf{x}\in\mathbb{C}^{n_{s}\times1}$ denote a unitary precoded
symbol vector transmitted by the source node. The $\mathrm{s-r}$
communication model can be accordingly expressed as,
\begin{equation}
\mathbf{y}_{\mathrm{r}}=\alpha_{\mathrm{sr}}\mathbf{H}_{\mathrm{sr}}\mathbf{x}+\mathbf{w}_{\mathrm{r}}\in\mathbb{C}^{n_{\mathrm{r}}\times1}.\label{eq:SR CM}
\end{equation}

The key idea of PF relaying is to extract and forward the DoFs of
the received signal vector $\mathbf{y}_{\mathrm{r}}$ via a QR-based
orthogonal projection \cite{Matrix_Computations}. Given that $\mathbf{H}_{\mathrm{sr}}$
is a pinhole channel, a single degree of freedom will be conveyed
by the relay to be used in the estimation of the transmit vector $\mathbf{x}$
at the destination.

Let $\mathbf{H}_{\mathrm{sr}}=\mathbf{Q}\mathbf{R}$ denote the QR
decomposition of $\mathbf{H}_{\mathrm{sr}}$, where $\mathbf{Q}\in\mathbb{C}^{n_{\mathrm{r}}\times n_{\mathrm{r}}}$
is a unitary matrix with $\mathbf{q}\in\mathbb{C}^{n_{\mathrm{r}}\times1}$
standing for its first column vector, and $\mathbf{R}\in\mathbb{C}^{n_{\mathrm{r}}\times n_{\mathrm{s}}}$
is an upper triangular matrix whose $(n_{\mathrm{r}}-1)$ bottom rows
consist entirely of zeros, i.e.,\vspace{-2mm}
\begin{equation}
\mathbf{R}=\left[\begin{array}{c}
\mathbf{h}_{\mathrm{r}}\\
\mathbf{0}_{(n_{\mathrm{r}}-1)\times n_{\mathrm{s}}}
\end{array}\right].\label{eq:R}
\end{equation}

The DoF $\hat{y}_{\mathrm{r}}$ is first obtained as 
\begin{equation}
\widetilde{y}_{\mathrm{r}}=\mathbf{q}^{\mathrm{H}}\mathbf{y}_{\mathrm{r}}=\alpha_{\mathrm{sr}}\mathbf{h}_{\mathrm{r}}\mathbf{x}+\mathbf{q}^{\mathrm{H}}\mathbf{w}_{\mathrm{r}}\in\mathbb{C},\label{eq:DoF}
\end{equation}
and is then normalized with a scaling factor $\alpha_{\mathrm{r}}=\left(\alpha_{\mathrm{sr}}^{2}\left\Vert \mathbf{h}_{\mathrm{r}}\right\Vert _{\mathrm{F}}^{2}+\sigma^{2}\right)^{\hspace{-1mm}-1/2}$
before being forwarded to the destination using only one relay antenna. The $\mathrm{r}$--$\mathrm{d}$ link is therefore a SISO Rayleigh channel whose
fading coefficient $h_{\mathrm{rd}}$ is rid of the antenna index, resulting in a simpler case
\begin{equation}
y_{\mathrm{d}}=\alpha_{\mathrm{rd}}\alpha_{\mathrm{r}}h_{\mathrm{rd}}\widetilde{y}_{\mathrm{r}}+w_{\mathrm{d}}\in\mathbb{C}.\label{eq:RD CM}
\end{equation}

\section{Performance Analysis}

\subsection{Instantaneous SNRs Characterization}

By invoking communication models \eqref{eq:DoF} and \eqref{eq:RD CM},
end-to-end SNR of the PF system in the mixed MIMO-pinhole/Rayleigh channel can be expressed similarly to a dual-hop
AF transmission \cite{Hasna}, i.e., 
\begin{equation}
\gamma_{\mathrm{srd}}=\frac{\gamma_{\mathrm{sr}}\gamma_{\mathrm{rd}}}{\gamma_{\mathrm{sr}}+\gamma_{\mathrm{rd}}+1},\label{eq:E2E SNR Compact}
\end{equation}
where the conditional terms $\gamma_{\mathrm{sr}}=\overline{\gamma}_{\mathrm{sr}}\left\Vert \mathbf{h}_{\mathrm{r}}\right\Vert _{\mathrm{F}}^{2}$
and $\gamma_{\mathrm{rd}}=\overline{\gamma}_{\mathrm{rd}}\left|h_{\mathrm{rd}}\right|^{2}$
represent the relay post-processing SNR and the destination receive
SNR, respectively.

To evaluate the PDF of $\gamma_{\mathrm{sr}}$, we consider the equality
$\mathbf{q}\mathbf{h}_{\mathrm{r}}=\mathbf{H}_{\mathrm{sr}}=\mathbf{g}_{\mathrm{r}}\mathbf{g}_{\mathrm{s}}^{\mathrm{H}}$
that stems from the aforementioned QR decomposition. Given that $\mathbf{q}$
is unitary, we infer that $\left\Vert \mathbf{h}_{\mathrm{r}}\right\Vert _{\mathrm{F}}=\left\Vert \mathbf{g}_{\mathrm{r}}\right\Vert _{\mathrm{F}}\left\Vert \mathbf{g}_{\mathrm{s}}\right\Vert _{\mathrm{F}}$,
and due to the statistical independence between $\mathbf{g}_{\mathrm{s}}$
and $\mathbf{g}_{\mathrm{r}}$, the PDF of $\left\Vert \mathbf{h}_{\mathrm{r}}\right\Vert _{\mathrm{F}}^{2}$ can be shown to be 
\vskip -0.2cm

{\small 
\begin{equation}
f_{\left\Vert \mathbf{h}_{\mathrm{r}}\right\Vert _{\mathrm{F}}^{2}}(\gamma)=\int_{0}^{+\infty}\frac{1}{\gamma_{\mathrm{r}}}f_{\left\Vert \mathbf{g}_{\mathrm{s}}\right\Vert _{\mathrm{F}}^{2}}\left(\frac{\gamma}{\gamma_{\mathrm{r}}}\right)f_{\left\Vert \mathbf{g}_{\mathrm{r}}\right\Vert _{\mathrm{F}}^{2}}(\gamma_{\mathrm{r}})\mathrm{d}\gamma_{\mathrm{r}}.\label{eq:hr2 pdf}
\end{equation}
}
By recalling that both $\mathbf{g}_{\mathrm{s}}$ and $\mathbf{g}_{\mathrm{r}}$
are Rayleigh fading vectors, we have $2\left\Vert \mathbf{g}_{\mathrm{i}}\right\Vert _{\mathrm{F}}^{2}\sim\mathcal{X}_{2n_{\mathrm{i}}}^{2},\mathrm{i}\hspace{-1mm}\in\hspace{-1mm}\left\{ \mathrm{s,r}\right\} $.
After some algebraic manipulations and by making use of \eqref{eq:hr2 pdf}
and \cite[Eq. (3.471.9)]{Table_of_Integrals}, we obtain the PDF of
$\gamma_{\mathrm{sr}}$ under the form
\vspace{-2mm}

{\small 
\begin{equation}
f_{\gamma_{\mathrm{sr}}}\left(\gamma\right)=\frac{2}{\Gamma(n_{\mathrm{s}})\Gamma(n_{\mathrm{r}})\overline{\gamma}_{\mathrm{sr}}}\hspace{-1mm}\left(\frac{\gamma}{\overline{\gamma}_{\mathrm{sr}}}\right)^{\hspace{-1.5mm}\frac{n_{\mathrm{s}}+n_{\mathrm{r}}}{2}-1}\hspace{-1.5mm}K_{n_{\mathrm{r}}-n_{\mathrm{s}}}\hspace{-1mm}\left(\hspace{-1mm}2\sqrt{\frac{\gamma}{\overline{\gamma}_{\mathrm{sr}}}}\right).\label{eq:SR SNR PDF}
\end{equation}
}

\vskip -0.2cm
The $\mathrm{r}$--$\mathrm{d}$ link is experiencing Rayleigh flat fading. Hence,
$\gamma_{\mathrm{rd}}$ is exponentially distributed with the probability
density function written as {\small $f_{\gamma_{\mathrm{rd}}}\left(\gamma\right)=(1/\overline{\gamma}_{\mathrm{rd}})\exp\left(-\gamma/\bar{\gamma}_{\mathrm{rd}}\right)$}.

\begin{figure*}[!t] 
\normalsize 
\setcounter{MYtempeqncnt}{\value{equation}} 
\setcounter{equation}{12} 
\vskip -0.4cm
{\small
\begin{equation} 
\label{eq:Pout}  
P_{\mathrm{out}}\left(\gamma_{\mathrm{th}}\right)=1\hspace{-0.5mm}-\hspace{-0.5mm}\frac{e^{-\frac{\gamma_{t\mathrm{h}}}{\overline{\gamma}_{\mathrm{rd}}}}}{\Gamma(n_{\mathrm{s}})\Gamma(n_{\mathrm{r}})}\sum_{k=0}^{+\infty}\frac{\left(-1\right)^{k}}{k!}\hspace{-0.5mm}\left(\frac{\gamma_{\mathrm{th}}+1}{\overline{\gamma}_{\mathrm{rd}}}\right)^{\hspace{-1mm}k}\sum_{l=0}^{+\infty}\frac{a_{k,l}}{l!}\left(\frac{\gamma_{\mathrm{th}}}{\overline{\gamma}_{\mathrm{sr}}}\right)^{\hspace{-1mm}\hspace{-0.5mm}k+l+1}\hspace{-2mm}\mathrm{G}_{1,3}^{3,0}\left(\frac{\gamma_{\mathrm{th}}}{\overline{\gamma}_{\mathrm{sr}}}\begin{array}{|c} 0\\ \hspace{-0.5mm}-1,\frac{\nu}{2}\hspace{-0.5mm}+\hspace{-0.5mm}\alpha\hspace{-0.5mm}-\hspace{-0.5mm}k\hspace{-0.5mm}-\hspace{-0.5mm}l,-\frac{\nu}{2}\hspace{-0.5mm}+\hspace{-0.5mm}\alpha\hspace{-0.5mm}-\hspace{-0.5mm}k\hspace{-0.5mm}-\hspace{-0.5mm}l \end{array}\hspace{-1mm}\right)
\end{equation}
}%
\setcounter{equation}{\value{MYtempeqncnt}} 
 \hrulefill 
\end{figure*}

\normalsize 
\setcounter{MYtempeqncnt}{\value{equation}} 
\setcounter{equation}{14} 

\begin{figure*}[!t] 
\normalsize 
\setcounter{MYtempeqncnt}{\value{equation}} 
\setcounter{equation}{14} 
\vspace{-6mm}
{\small
\begin{align} 
\overline{C}_{\mathrm{srd}}=&\frac{1}{2\ln\left(2\right)\Gamma(n_{\mathrm{s}})\Gamma(n_{\mathrm{r}})}\sum_{k=0}^{+\infty}\frac{\left(-1\right)^{k}}{k!}\sum_{n=0}^{k}\binom{k}{n}\sum_{l=0}^{+\infty}\frac{a_{k,l}}{l!}\frac{\overline{\gamma}_{\mathrm{rd}}^{l+n+1}}{\overline{\gamma}_{\mathrm{sr}}^{k+l+1}}\nonumber\\
&\qquad\times\left[\mathrm{G}_{1,0:2,2:1,3}^{1,0:1,2:3,0}\left(\hspace{-1mm}\begin{array}{c|} \overline{\gamma}_{\mathrm{rd}},\frac{\overline{\gamma}_{\mathrm{rd}}}{\overline{\gamma}_{\mathrm{sr}}}\end{array}\begin{array}{c} k+l+n+2\\ \_ \end{array}\begin{array}{|c} 1,1\\ 1,0 \end{array}\begin{array}{|c} 0\\ -1,\frac{\nu}{2}+\alpha-k-l,-\frac{\nu}{2}+\alpha-k-l \end{array}\hspace{-1mm}\right)\right.\nonumber\\
&\qquad\quad\left.-\mathrm{G}_{1,0:2,2:2,4}^{1,0:1,2:3,1}\left(\hspace{-1mm}\begin{array}{c|} \overline{\gamma}_{\mathrm{rd}},\frac{\overline{\gamma}_{\mathrm{rd}}}{\overline{\gamma}_{\mathrm{sr}}}\end{array}\begin{array}{c} k+l+n+1\\ \_ \end{array}\begin{array}{|c} 1,1\\ 1,0 \end{array}\begin{array}{|c} -(k+l+n+1),0\\ -1,\frac{\nu}{2}+\alpha-k-l,-\frac{\nu}{2}+\alpha-k-l,-(k+l+n) \end{array}\hspace{-1mm}\right)\right]\label{eq:Csrd}
\end{align}
}%
\vspace{-2.1mm}
\setcounter{equation}{\value{MYtempeqncnt}} 
\hrulefill 
\end{figure*}

\begin{figure*}[!t] 
\normalsize 
\setcounter{MYtempeqncnt}{\value{equation}} 
\setcounter{equation}{15} 
\vspace{-3mm}
{\small 
\[-\mathrm{Res}\left[{\frac{\Gamma(-1-\hspace{-1mm}s)\Gamma(\frac{\nu}{2}+\hspace{-1mm}\alpha-\hspace{-1mm}k-\hspace{-1mm}l-\hspace{-1mm}s)\Gamma(-\frac{\nu}{2}+\hspace{-1mm}\alpha-\hspace{-1mm}k-\hspace{-1mm}l-\hspace{-1mm}s)}{\Gamma(-s)}\left(\hspace{-0.5mm}\frac{\gamma_{\mathrm{th}}}{\overline{\gamma}_{\mathrm{sr}}}\hspace{-0.5mm}\right)^{\hspace{-1mm}s}}\hspace{-1mm},-1\right]=\Gamma\left(\frac{\nu}{2}+\hspace{-1mm}\alpha-\hspace{-1mm}k-\hspace{-1mm}l+\hspace{-1mm}1\right)\Gamma\left(-\frac{\nu}{2}+\hspace{-1mm}\alpha-\hspace{-1mm}k-\hspace{-1mm}l+\hspace{-1mm}1\right)\left(\hspace{-0.5mm}\frac{\gamma_{\mathrm{th}}}{\overline{\gamma}_{\mathrm{sr}}}\hspace{-0.5mm}\right)^{\hspace{-1mm}-1} \label{eq:Residue1}\]
}
\vspace{-7mm}
\setcounter{equation}{\value{MYtempeqncnt}} 
\hrulefill 
\end{figure*}

\begin{figure*}[!t] 
\normalsize 
\setcounter{MYtempeqncnt}{\value{equation}} 
\setcounter{equation}{18} 
\vspace*{4mm}
{\small
\[
\mathrm{Res}\left[\underbrace{\frac{\Gamma^{2}\left(1-s\right)\Gamma^{3}\left(s\right)\Gamma\left(n_{\mathrm{r}}+s\right)\Gamma\left(n_{\mathrm{s}}\hspace{-0.5mm}+\hspace{-0.5mm}s\right)}{\Gamma^{2}\left(1+s\right)\Gamma\left(-s\right)}\overline{\gamma}_{\mathrm{sr}}^{s}}_{\phi(s)},0\right]=\frac{1}{2}\lim_{s\rightarrow0}\frac{\mathrm{d}^{2}}{\mathrm{d}s^{2}}\hspace{-0.5mm}\left[s^{3}\phi(s)\right]=\Gamma(n_{\mathrm{s}})\Gamma(n_{\mathrm{r}})\left[\ln\left(\overline{\gamma}_{\mathrm{sr}}\right)\hspace{-0.5mm}+\hspace{-0.5mm}\Psi^{(0)}\left(n_{\mathrm{s}}\right)\hspace{-0.5mm}+\hspace{-0.5mm}\Psi^{(0)}\left(n_{\mathrm{r}}\right)\hspace{-0.5mm}\right]\label{eq:Residue2}\]
}
\vspace{-6mm}
\setcounter{equation}{\value{MYtempeqncnt}} 
\hrulefill 
\end{figure*}

\setcounter{equation}{8}

\vspace{-2mm}
\subsection{Outage Probability}

In noise-limited transmissions, quality of service (QoS) is ensured by keeping the instantaneous end-to-end SNR above a threshold $\gamma_{\mathrm{th}}$. The probability of outage in our relaying setup is expressed as 
\vskip -0.3cm

{\small
\begin{equation}
P_{\mathrm{out}}=\mathbf{\Pr}\left[\gamma_{\mathrm{srd}}<\gamma_{\mathrm{th}}\right]=\mathbf{\Pr}\left[\frac{\gamma_{\mathrm{sr}}\gamma_{\mathrm{rd}}}{\gamma_{\mathrm{sr}}+\gamma_{\mathrm{rd}}+1}<\gamma_{\mathrm{th}}\right],
\end{equation}
}
\vskip -0.2cm
\noindent which is actually the cumulative distribution function (CDF) of SNR $\gamma_{\mathrm{srd}}$. Marginalization over $\gamma_{\mathrm{sr}}$ yields
\vskip -0.25cm

{\small
\begin{equation}
P_{\mathrm{out}}\left(\gamma_{\mathrm{th}}\right)=1-\int_{0}^{+\infty}\widetilde{F}_{\gamma_{\mathrm{\mathrm{rd}}}}\left[\gamma_{\mathrm{th}}+\frac{\gamma_{\mathrm{th}}^{2}+\gamma_{\mathrm{th}}}{\gamma}\right]f_{\gamma_{\mathrm{sr}}}\left(\gamma\right)\mathrm{d}\gamma,
\end{equation}
}

\vskip -0.25cm
\noindent where $\widetilde{F}_{\gamma_{\mathrm{rd}}}\left(\cdot\right)$ is the complementary CDF (CCDF) of
$\gamma_{\mathrm{rd}}$, given by $\exp\left(-\gamma/\bar{\gamma}_{\mathrm{rd}}\right)$.
By plugging \eqref{eq:SR SNR PDF} into the above integral and making
the change $u=1+\gamma/\gamma_\mathrm{th}$ as well as a Taylor expansion of an exponential term, we infer that

{\small
\begin{equation}\label{eq:Pout1}
P_{\mathrm{out}}\left(\gamma_{\mathrm{th}}\right)=1-2\left(\frac{\gamma_{\mathrm{th}}}{\overline{\gamma}_{\mathrm{sr}}}\right)^{\hspace{-0.5mm}\alpha+1}\hspace{-0.5mm}\frac{e^{-\frac{\gamma_{\mathrm{th}}}{\overline{\gamma}_{\mathrm{rd}}}}}{\Gamma(n_{\mathrm{s}})\Gamma(n_{\mathrm{r}})}\times\mathcal{I}, 
\end{equation}
}
with the term $\mathcal{I}$ given by
{\small
\begin{eqnarray}
\mathcal{I}=\sum_{k=0}^{+\infty}\frac{\left(-1\right)^{k}}{k!}\left(\frac{\gamma_{\mathrm{th}}+1}{\overline{\gamma}_{\mathrm{rd}}}\right)^{k}\sum_{l=0}^{+\infty}\frac{a_{k,l}}{l!}\,\,\,\,\,\,\,\,\,\,\,\,\,\,\,\,\,\,\,\,\,\,\,\,\,\,\,\,\,\,\,\,\,\,\,\,\,\,\,\,\,\nonumber \\
\times\int_{1}^{+\infty}u^{\alpha-k-l}K_{\nu}\left(2\sqrt{\frac{\gamma_{\mathrm{th}}}{\overline{\gamma}_{\mathrm{sr}}}u}\right)\mathrm{d}u,\label{eq:Iout}
\end{eqnarray}
}
where $\alpha=(n_{\mathrm{s}}+n_{\mathrm{r}})/2-1$, $\nu=n_{\mathrm{r}}-n_{\mathrm{s}}$, and
$a_{k,l}=\Gamma\left(k+l\right)/\Gamma\left(k\right)$
with the particular case $a_{0,0}=1$. Then, by combining \eqref{eq:Pout1}
and \eqref{eq:Iout} and using \cite[Eq. (6.592.4)]{Table_of_Integrals},
an exact expression of $P_{\mathrm{out}}$ is obtained after
some simplifications as shown in \eqref{eq:Pout} on top of the next~page.

\setcounter{equation}{13}

\subsection{Ergodic Capacity}

Unlike the approximation in~\cite{Waqar}, the end-to-end ergodic capacity of the dual-hop PF system under consideration can be written as
\begin{equation}
\overline{C}_{\mathrm{srd}}=\frac{1}{2}\int_{0}^{+\infty}\log_{2}\!\left(1+\gamma\right)f_{\gamma_{\mathrm{srd}}}\!\left(\gamma\right)\mathrm{d}\gamma,\label{eq:Ergodic Capacity}
\end{equation}
where $f_{\gamma_{\mathrm{srd}}}$ is the PDF of $\gamma_{\mathrm{srd}}$
that is computed by firstly expanding the power $\left(\gamma_{\mathrm{th}}+1\right)^{k}$
in \eqref{eq:Pout} into a finite sum using the Binomial theorem.
The resulting function is then differentiated with respect to $\gamma_{\mathrm{th}}$
via \cite[Eq. (5)]{Wolfram}. By rewriting the elementary functions involved in the obtained PDF  as Meijer G-functions \cite[Eq. (11)]{Elementary_MeijerG_Equivalent}, i.e.,
$\gamma^{p}e^{-\frac{\gamma}{\overline{\gamma}_{\mathrm{rd}}}}=\overline{\gamma}_{\mathrm{rd}}^{p}\mathrm{G}_{0,1}^{1,0}\!\left(\!\displaystyle{\frac{\gamma}{\overline{\gamma}_{\mathrm{rd}}}}\begin{array}{|c}
-\\
p
\end{array}\!\right)$ and $\ln\left(1{+}\gamma\right){=}\mathrm{G}_{2,2}^{1,2}\!\left(\!
\gamma~ \begin{array}{|c}
1,1\\
1,0
\end{array}\!\right)$, the ergodic capacity is expressed in terms of integrals of the product
of three Meijer G-functions whose expressions are given in terms of
the Bivariate Meijer G-function according to \cite[Eq. (12)]{Bivariate_Meijer_G}
as shown in \eqref{eq:Csrd}.

\setcounter{equation}{16}

\section{Asymptotic Behavior}

To highlight the effect of channel parameters on both the outage probability
and the ergodic capacity, we study their asymptotic behaviors. Invoking
\cite[Theorem 1.7 and Theorem 1.11]{Kilbas}, expansions of the Mellin-Barnes
integrals involved in the Meijer-G and bivariate Meijer-G functions
can be derived by evaluating the residue of the corresponding integrands
at the pole closest to the contour; the minimum pole on the right
$p_\textrm{min}^{-}$ for small Meijer-G arguments and the maximum pole on
the left $p_\textrm{max}^{+}$ for large ones, as depicted in Fig. 1. Moreover,
the \emph{Inside-Outside} theorem \cite{Inside-Outside} states that
the obtained result is further multiplied by $-1$ in the case of a clockwise-oriented contour (i.e., for small arguments).

\vspace{-0.5mm}
\subsection{Asymptotic Outage Probability}

We study the asymptotic behavior of the outage probability for a low SNR threshold $\gamma_{\mathrm{th}}$. By keeping low order terms in \eqref{eq:Pout},
i.e., $k+l\leq1$, and given that $\alpha+\nu/2=n_{\mathrm{r}}-1\geq1$
and $\alpha-\nu/2=n_{\mathrm{s}}-1\geq1$, we have $\pm\nu/2+\alpha-k-l\geq0$.
Therefore, we evaluate the residue at $-1$ (that is the smallest pole) as shown in \eqref{eq:Residue1}. Replacing the exponential function with its first order expansion near zero, $\exp(-\frac{\gamma_{\mathrm{th}}}{\overline{\gamma}_{\mathrm{rd}}})\thickapprox1-\frac{\gamma_{\mathrm{th}}}{\overline{\gamma}_{\mathrm{rd}}}$, yields the following asymptotic expression:

\vspace{-3mm}
\begin{equation}
P_{\mathrm{out}}\left(\gamma_{\mathrm{th}}\right)=\left(1+\frac{1}{\left(n_{\mathrm{s}}-1\right)\left(n_{\mathrm{r}}-1\right)\overline{\gamma}_{\mathrm{sr}}}\right)\frac{\gamma_{\mathrm{th}}}{\overline{\gamma}_{\mathrm{rd}}}+o\left(\gamma_{\mathrm{th}}\right).
\end{equation}

\begin{figure}
\vspace{-1mm}
\begin{centering}
\includegraphics[scale=0.55]{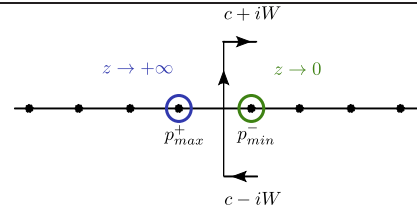}
\par\end{centering}
\caption{Complex contour of the Mellin-Barnes integral of argument $z$. $W$~is~set to a large
value.}
\vspace{-0.35cm}
\end{figure}
\vspace{-6mm}

\subsection{Asymptotic Ergodic Capacity}

Based on \eqref{eq:Csrd}, the asymptotic behavior of the ergodic
capacity is derived for different scenarios of the balance parameter $\beta=\frac{\overline{\gamma}_{\mathrm{rd}}}{\overline{\gamma}_{\mathrm{sr}}}$
and the SNR $\overline{\gamma}_{\mathrm{sr}}$ as summarized in Table I. Let $s$ and $t$ denote the integration variables in the bivariate Meijer-G function. In the case $\beta\rightarrow+\infty$, we evaluate the residue of the first and second bivariate Meijer-G terms in \eqref{eq:Csrd} at the highest poles on the left of the contour, i.e., $t=-(k+l+n+2+s)$ and $t=-(k+l+n+1+s)$, respectively. Keeping only 0-th orders on $1/\beta$ results in the expression (18). Expression (20) is inferred by computing the residue of the integrand of the Meijer-G term in (18) at $s=0$ as shown in \eqref{eq:Residue2}. The remaining cases are obtained using the same approach.

\vskip -2mm
\begin{table}[!h]
\centering
\caption{Ergodic capacity asymptotic expressions}\vspace{-2mm}
\begin{tabular}{|p{2.2cm}|p{5.7cm}|}
\hline 
Scenario & \hspace{1.8cm} Asymptotic $\overline{C}_{srd}$\\
\hline 
\hline 
$\beta\rightarrow+\infty$ & $\begin{array} {lcl} \frac{1}{2\ln(2)\Gamma(n_{\mathrm{s}})\Gamma(n_{\mathrm{r}})}&& \\ \times \mathrm{G}_{6,4}^{2,5}\left(\overline{\gamma}_{\mathrm{sr}}\begin{array}{|c}
1,1,1,1-n_{\mathrm{r}},1-n_{\mathrm{s}},0\\
1,1,0,0
\end{array}\hspace{-1mm}\right) \hspace{1mm}\stepcounter{equation}\thetag{\theequation}&& \end{array}$\\
\hline 
\setcounter{equation}{19}
$\beta,\overline{\gamma}_{\mathrm{sr}}\rightarrow+\infty$ & $\frac{1}{2\ln(2)}\left[\ln\left(\overline{\gamma}_{\mathrm{sr}}\right)+\psi^{(0)}\left(n_{\mathrm{s}}\right)+\psi^{(0)}\left(n_{\mathrm{r}}\right)\right]$\hspace{1mm}\stepcounter{equation}\thetag{\theequation}\\
\hline
$\beta\rightarrow0$ or $\overline{\gamma}_{\mathrm{sr}}\rightarrow0$ & \hspace{2.5cm}$0$\\
\hline
\end{tabular}

\vskip -0.25cm
\end{table}
\vspace{-0.25cm} 
\section{Numerical Results}

In this section, we present a few numerical results to illustrate the theoretical analysis. For different antenna and SNR setups, Fig. 2 and 3 show the exact and asymptotic results of both the end-to-end outage probability and the ergodic capacity, respectively. Throughout our numerical experiments, we found out that regardless of the average SNRs and antennas settings, accurate analytical curves can be obtained by truncating the infinite sums at $K=50$ and $L=5$ terms. The exact match with Monte-Carlo simulation results confirms the precision of the theoretical analysis. As the PF scheme is a variant of AF, also operating at the signal-level, per antenna CSI-assisted AF simulations are provided for comparison. The bivariate Meijer G-function with automated contour---presented in the Appendix---was developed to enable the numerical evaluation of \eqref{eq:Csrd} in MATLAB environments. For the sake of precision, we note that the contour length $W$ should be increased (e.g., 10 and more) for high arguments.

\begin{figure}
\vskip -0.5cm
\includegraphics[scale=0.6]{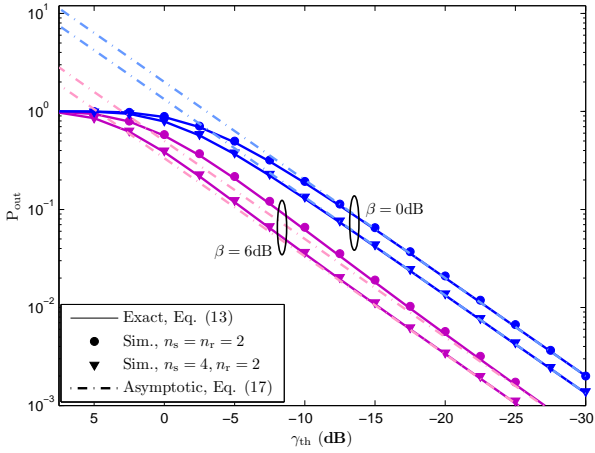}
\vspace{-0.4cm} 
\caption{End-to-end outage probability versus $\gamma_{\mathrm{th}}$ for $\overline{\gamma}_{\mathrm{sr}}=0$
dB.}
\vskip -0.2cm
\end{figure}

\vskip -0.3cm

\section{Conclusion}

In this letter, we have presented a performance evaluation
of dual-hop PF systems over the practical mixed MIMO-pinhole/Rayleigh channel.
For numerical evaluation purposes, we have proposed a novel and fast
MATLAB implementation of the bivariate Meijer-G function. Exact and asymptotic
results are in total agreement with Monte-Carlo simulations,
and can be used by system designers to define SNR thresholds for switching
between PF and other relaying schemes in pinhole conditions. 

\vspace{-0.4cm} 
\appendix{\noun{~~~~Bivariate Meijer G-Function's Matlab Code}}

\lstset{
language=Matlab,    
keywords={break,case,catch,continue,else,elseif,end,for,function,global,if,otherwise,persistent,return,switch,try,while},    basicstyle=\scriptsize\ttfamily, 
keywordstyle=\color{blue},    
commentstyle=\color{dkgreen}, 
stringstyle=\color{Purple},
numberstyle=\tiny\color{gray},    
stepnumber=1,    
numbersep=10pt,    
backgroundcolor=\color{white},    
tabsize=4,    
showspaces=false,    
showstringspaces=false}
\begin{lstlisting} 
function out = Bivariate_Meijer_G(am1, ap1, bn1, bq1, cm2, ...
                      cp2, dn2, dq2, em3, ep3, fn3, fq3, x, y)

%***** Integrand definition *****
F = @(s,t)(GammaProd(am1,s+t).* GammaProd(1-cm2,s) ...
 .* GammaProd(dn2,-s) .* GammaProd(1-em3,t) ...
.* GammaProd(fn3,-t).* (x.^s) .* (y.^t)) ...
./(GammaProd(1-ap1,-(s+t)).* GammaProd(bq1,s+t) ...
.* GammaProd(cp2,-s) .* GammaProd(1-dq2,s) ...
.* GammaProd(ep3,-t) .* GammaProd(1-fq3,t));
%***** Contour definition *****
Sups = min(dn2); Infs = -max(1-cm2); % cs 
cs = (Sups + Infs)/2;% s between Sups and Infs 
Supt = min(fn3); Inft = max([-am1-cs em3-1]);% t>-am1-s,s=cs 
ct = Supt - ((Supt - Inft)/10);% t between Supt and Inft 
W = 10; % W
%***** Bivariate Meijer G *****
out = (-1/(2*pi)^2)*quad2d(F,cs-j*W,cs+j*W,ct-j*W,ct+j*W,...
'AbsTol',10^-5,'RelTol',10^-5,'MaxFunEvals',2000,...
'Singular',true); %Increase MaxFunEvals for higher W
%***** GammaProd subfunction *****
	function output = GammaProd(p,z) 
	[pp zz] = meshgrid(p,z); 
		if (isempty(p)) output = ones(size(z));
		else output = reshape(prod(gamma(pp+zz),2),size(z));
		end
	end
% The gamma function here is the complex gamma, available in
% www.mathworks.com/matlabcentral/fileexchange/3572-gamma
end

\end{lstlisting}\vskip -0.7cm

\begin{figure}
\vskip -0.5cm
\includegraphics[scale=0.6]{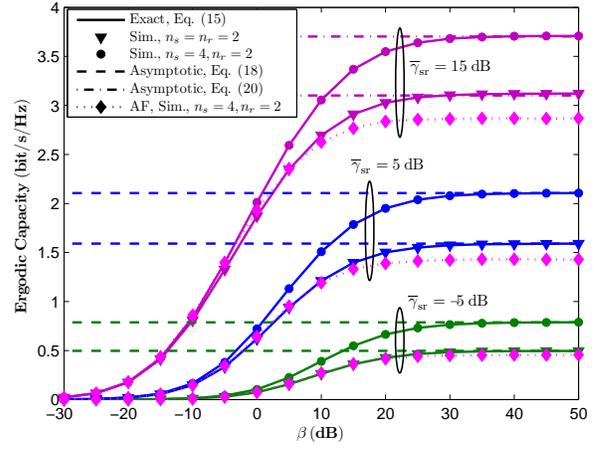}
\vspace{-0.3cm} 
\caption{End-to-end ergodic capacity versus $\beta$ for different SNR $\overline{\gamma}_{\mathrm{sr}}$ and~antennas configurations.}
\vskip 0.2cm
\end{figure}


\begin{thebibliography}{10}

\bibitem[1]{MIMO_Outdoor}D. Gesbert, H. Bölcskei, D. A. Gore, and A. J. Paulraj, ``Outdoor MIMO Wireless Channels: Models and Performance Prediction,'' \emph{IEEE Trans. Commun.}, vol. 50, no. 12, pp. 1926-1934,
Dec. 2002.

\bibitem[2]{Long_Distance}J. M. Vella, S. Zammit, ``Performance improvement of long distance MIMO links using cross polarized antennas,'' in \emph{15th IEEE Mediterranean Electrotechnical Conference (MELECON)}, Valletta, Malta, 26-28 Apr. 2010, pp. 1287-1292.

\bibitem[3]{MRN}A. O. Laiyemo, P. Pirinen, M. Latva-aho, J. Vihriala, V. Van Phan, ``Impact of LTE precoding for fixed and adaptive rank transmission in moving relay node system,'' in \emph{ITS Telecommunications (ITST)}, 5-7 Nov. 2013, pp. 250-254.

\bibitem[4]{Multihop_DMT}S. Yang and J.-C. Belfiore, ``Diversity
of MIMO Multihop Relay Channels--Part I: Amplify-and-Forward,''\textit{
IEEE Trans. Inform. Theory, Submitted, available at http://arxiv.org/abs/0708.0386}

\bibitem[5]{Keyhole1}A. Firag, H. A. Suraweera, P. J. Smith and C. Yuen, ``Dual-hop MIMO amplify-and-forward relay channel capacity with keyhole effect,'' \emph{IEEE Commun. Letters}, vol. 15, no. 10, pp. 1050-1052, Oct. 2011.

\bibitem[6]{Keyhole2}T. Q. Duong, H. A. Suraweera, T. A. Tsiftsis, H. Zepernick, A. Nallanathan, ``OSTBC Transmission in MIMO AF Relay Systems with Keyhole and Spatial Correlation Effects,'' in \emph{IEEE International Conference on Communications (ICC)}, Kyoto, Japan, 5-9 Jun. 2011, pp.1-6.

\bibitem[7]{Bivariate_Meijer_G_def}B. L. Sharma and R. F. A. Abiodun, ``Generating function for generalized function of
two variables,'' \emph{in Proc. American Mathematical Society,} vol. 46, no. 1, pp. 69-72, Oct. 1974.

\bibitem[8]{Table_of_Integrals}I. S. Gradshteyn and I. M. Ryzhik,
\emph{Table of Integrals, Series, and Products}, 7th ed., Academic Press, 2007.

\bibitem[9]{BER_Bivariate_Meijer_G}I. S. Ansari, S. Al-Ahmadi, F. Yilmaz, M.-S. Alouini, and H. Yanikomeroglu, ``A
new formula for the BER of binary modulations with dual-branch selection
over Generalized-K composite fading channels,'' \emph{IEEE Trans. Commun.,} vol. 59, no. 10, pp. 2654-2658,
Oct. 2011.

\bibitem[10]{Matrix_Computations}G. H. Golub and C. F. Van Loan,
\textit{Matrix Computations (Third Edition)}, The John Hopkins University Press, 1996.

\bibitem[11]{Hasna}M. O. Hasna and M.-S. Alouini, ``End-to-end performance of transmission systems with relays over Rayleigh-fading channels,'' \emph{IEEE Trans. Wireless Commun.,} vol. 2, no. 6, pp. 1126-1131, Nov. 2003.

\bibitem[12]{Waqar}O. Waqar, M. Ghogho, D. McLernon, ``Performance analysis of dual-hop variable gain relay networks over Generalized-K fading channels,'' in \emph{IEEE Eleventh International Workshop on Signal Processing Advances in Wireless Communications (SPAWC)}, 20-23 Jun. 2010, pp.1-5.

\bibitem[13]{Wolfram}http://functions.wolfram.com/HypergeometricFunctions/MeijerG/20/01/01/

\bibitem[14]{Elementary_MeijerG_Equivalent}V. S. Adamchik and O. S.
Marichev, ``The algorithm for calculating integrals
of hypergeometric type functions and its realization in REDUCE system,'' \emph{in Proc. International Symposium on Symbolic and Algebraic Computation},
ACM, Academic Press, pp 212-224, 1990.

\bibitem[15]{Bivariate_Meijer_G}S. C. Gupta, ``Integrals involving products of G-functions,'' \emph{Proceedings of the National Academy of Sciences,} India, vol. 39(A), no. II, 1969

\bibitem[16]{Kilbas}A. Kilbas, \emph{H-Transforms: Theory and Applications.
Analytical Methods and Special Functions}, Taylor \& Francis, 2004.

\bibitem[17]{Inside-Outside}T. Rowland and E. W. Weisstein, ``Inside-Outside
Theorem''. From MathWorld -- A Wolfram Web Resource. http://mathworld.wolfram.com/Inside-OutsideTheorem.html.

%
%
%

\end{thebibliography}
\end{document}